\documentclass[journal]{IEEEtran}

\usepackage[T1]{fontenc}
\usepackage[utf8]{inputenc}
\usepackage{cite}
\usepackage{amsmath,amssymb,amsfonts}
\usepackage{mathtools}
\usepackage{bm}
\usepackage{graphicx}
\usepackage{booktabs}
\usepackage{multirow}
\usepackage{array}
\usepackage{xcolor}
\usepackage{url}
\usepackage{hyperref}
\usepackage{enumitem}

\usepackage{algorithm}
\usepackage{algpseudocode}

\hypersetup{hidelinks}

\newcommand{\me}{\mathsf{e}}
\newcommand{\mz}{\mathsf{z}}

\newcommand{\sfp}{\mathsf{p}}

\newcommand{\jmathi}{\mathrm{j}}

\newcommand{\dab}{DAB+~}
\newcommand{\dqpsk}{$\frac{\pi}{4}$-QPSK~}

\newtheorem{proposition}{\bf Proposition}

\newtheorem{remark}{\bf Remark}

\title{\huge Posterior-Aware Differential Channel Tracking for Reliable Single-Stream DAB+ Passive Radar}

\author{Kai~Wu,~\IEEEmembership{Member,~IEEE}, Brendan J. Hall, Zhongqin Wang,~\IEEEmembership{Member,~IEEE},  \\
J. Andrew Zhang,~\IEEEmembership{Senior Member,~IEEE}, and Y. Jay Guo,~\IEEEmembership{Life Fellow,~IEEE}

\thanks{K. Wu, B. Hall, Z. Wang, J. A. Zhang and Y. J. Guo are with the Global Big Data Technologies Centre (GBDTC), University of Technology Sydney (UTS), Sydney, NSW 2007, Australia. E-mail: kai.wu@uts.edu.au; brendan.j.hall@student.uts.edu.au; zhongqin.wang@uts.edu.au; andrew.zhang@uts.edu.au; jay.guo@uts.edu.au.}
}

\begin{document}

\maketitle

\begin{abstract}
Digital audio broadcasting plus (DAB+) is an attractive illuminator for passive radar because it provides persistent, high-power, and geographically widespread very high frequency (VHF) orthogonal frequency-division multiplexing (OFDM) signals. A channel state information (CSI) sensing approach can convert a single received DAB+ stream into a CSI sequence for radar sensing, avoiding the need for a separately received reference signal in conventional passive radars. However, CSI estimation in DAB+ is challenging due to the differentially encoded communication symbols across time. A wrong symbol transition estimation leads to a persistent multiplicative error in the sequential CSI sequence within a DAB+ frame. This paper formulates single-stream DAB+ passive radar as a posterior-probability-aware differential CSI tracking problem. The proposed method uses the previously tracked CSI as a channel prior, performs prediction-aided maximum a posteriori detection of current symbol, converts posterior transition reliability into observation uncertainty, and applies linear minimum mean squared error fusion to obtain a stable tracking CSI. A reliability-informed CSI fusion strategy is also introduced to preserve weak target information. Theoretical analysis is provided, showing guaranteed performance again in symbol and CSI estimation. 
Simulation results show that 
the proposed method can reduce CSI estimation error by over 15~dB compared with prior art. 
It also improves median target-to-background ratio by more than 11~dB in random fading scenes.
Experiments in Sydney, Australia demonstrate improved range-Doppler maps for commercial aircraft sensing.
\end{abstract}


\section{Introduction}
\label{sec:intro}

Passive radar exploits non-cooperative transmitters as illuminators of opportunity and hence enables surveillance without dedicated radar emissions or additional spectrum. Among such illuminators, broadcast waveforms are particularly attractive because they are persistent, high-power, and geographically widespread \cite{kuschel2019tutorial,colone2022passive,taes_dab_localization2023}. Poullin showed that digital broadcasters using coded orthogonal frequency division multiplexing (OFDM), including digital audio broadcasting (DAB) and digital video broadcasting (DVB), can support passive detection \cite{Poullin2005PassiveDetection}. Coleman and Yardley later demonstrated passive bistatic radar based on target illumination by DAB transmitters \cite{Coleman2008DABPBR}, while Coleman \emph{et al.} reported a practical bistatic passive radar system using DAB and digital radio mondiale (DRM) illuminators \cite{Coleman2008PracticalDABDRM}. These early studies established DAB-family broadcasts as credible very high frequency (VHF) illuminators for passive coherent location (PCL) \cite{RecentProgressPCL2008}.

Subsequent work has broadened DAB-based passive radar from proof-of-concept detection to more complete sensing and surveillance systems. 
Schr\"oder \emph{et al.} investigated multiband experimental PCL operation with broadcast illuminators \cite{Schroeder2010MultibandPCL}. Reference signal reconstruction was shown to improve DAB-based passive bistatic radar detection in \cite{OHagan2010SignalReconstruction}. Beyond target detection, DAB/DVB illumination has been used for multistatic tracking \cite{Daun2012Tracking}, multi-illuminator system design and performance evaluation \cite{Edrich2014MultiIlluminator}, micro-UAV detection \cite{MicroUAV2017DAB}, DAB signal preprocessing for PCL \cite{Sensors2022DABPreprocessing}, and aircraft propeller-signature observation \cite{Propeller2025DAB}.

Most of these studies follow a reference channel-enabled PCL paradigm. A clean representation of the transmitted waveform is separately received or reconstructed, strong direct-path and clutter components are suppressed, and target signatures are extracted through delay-Doppler correlation. This paradigm is effective, but it treats the communication waveform primarily as an external illuminator. It does not fully exploit the waveform structure of broadcast OFDM signal. The importance of such structure has been recognized in related passive radar studies. Palmer \emph{et al.} analyzed DVB-T passive-radar signal processing and its waveform-dependent artifacts \cite{Palmer2013DVBTPassiveRadar}, while Evers and Jackson showed that communication-waveform features can have a direct impact on passive-radar ambiguity behavior \cite{Evers2015CrossAmbiguity}.

A complementary line of work has exploited OFDM channel estimates rather than relying solely on time-domain cross-ambiguity processing. Berger \emph{et al.} showed that OFDM block channel estimates can be used directly for passive radar target detection through Fourier analysis across successive OFDM blocks \cite{Berger2010OFDMPassiveRadar}. Gassier \emph{et al.} developed a frequency-domain framework in which disturbance cancellation and target detection are unified for OFDM passive radar \cite{gassier2016unifying}, and Chabriel and Barr\`ere extended this channel domain view to spatial-diversity receivers \cite{chabriel2017adaptive}. From a parameter-estimation perspective, Zheng and Wang formulated delay-Doppler estimation from OFDM passive-radar measurements as a super-resolution sparse recovery problem while explicitly accounting for demodulation errors \cite{zheng2017super}. Moreover, Lyu \emph{et al.} analyzed how inter-subcarrier interference affects weak-target detection in OFDM passive radar \cite{Lyu2024CEMAnalysis}. 

Recently, using OFDM channel estimation for sensing has been extensively studied for the highly popular integrated sensing and communications (ISAC). In cellular systems, channel state information (CSI) has been used for moving-target parameter estimation\footnote{As widely used in OFDM communication systems, e.g., cellular and WiFi, the term CSI is the channel frequency response over orthogonal subcarriers.} \cite{Rahman2020PMNFramework,Wu2022FlexibleSensing,Chen2023MUSICTWC} and environmental sensing \cite{KaiWu2025HumanEnvironmentalSensing}. Uplink and bistatic CSI sensing have further motivated methods that explicitly address clock asynchronism between spatially separated transceivers \cite{Ni2021UplinkAsynchronousPMN,KaiWu2023CSIRatioUplink,Hu2024CSIRatioBounds,KaiWu2024Asynchronism}. In parallel, WiFi sensing has shown that CSI variations can support device-free localization, activity recognition, and other environment-aware applications using commodity WLAN hardware \cite{Yousefi2017WiFiCSISurvey,Ma2019WiFiCSISurvey,Tan2022CommodityWiFiTenYears}. 
Broader treatments of CSI sensing in ISAC can be found in recent surveys and tutorials \cite{Zhang2021JCRSOverview,Zhang2022PMNSurvey,Liu2022ISACJSAC,Koivunen2024MulticarrierISAC,GonzalezPrelcic2024ISACRevolution,Du2025IEEE80211bfOverview}.

The CSI sensing viewpoint is attractive for DAB-based passive radar because it suggests that a single received OFDM stream can be converted into a channel sequence for radar processing, without needing an extra reference receiver for creating a local sensing signal template \cite{DAB_ofdm_sensing}. 
However, 
the CSI estimation using DAB+ signals is fundamentally different from that in the ISAC literature mentioned above. In cellular and WiFi systems, CSI is obtained from known pilots, reference signals, or channel-sounding procedures \cite{KaiWu2025HumanEnvironmentalSensing}. 
In DAB+, by contrast, there is no pilot or reference for channel estimation, and the useful OFDM symbols are differentially encoded over time on each active subcarrier \cite{ETSIEN3004012017}.

This differential structure creates the central challenge addressed in this paper. The unknown DAB+ symbol and the unknown CSI are multiplicatively coupled on each active subcarrier. 
The recent DAB+ passive radar work in \cite{DAB_ofdm_sensing} first estimates the differential QPSK (DQPSK) transition, reconstructs the absolute transmitted symbol, and then divides the reconstructed symbol out of the received signal on a subcarrier to obtain a CSI estimate. 
However, for radar sensing, such an open-loop differential decoding procedure is fragile. A wrong DQPSK transition decision on one subcarrier is not a single isolated outlier. It rotates all subsequent reconstructed symbols on that subcarrier in the same frame. The multiplicative error contaminates the CSI sequence, possibly over many symbols, which can substantially degrade radar performance.

{This paper develops a robust CSI estimation scheme for DAB+ passive radar by introducing 
a posterior-probability-aware differential CSI tracking framework.}
The proposed framework uses the previous tracking CSI as a channel prior for current-symbol reconstruction, quantifies the reliability of the resulting DQPSK decision, and uses this reliability to control how much instantaneous channel innovation is admitted for minimizing channel estimation error. 
To the best of our knowledge, this is the first systematic treatment of reliable CSI estimation
for DAB-based passive radar. 
The main contributions are summarized as follows.

\begin{itemize}
    \item We formulate single-stream DAB-based passive radar as a novel differential CSI tracking problem. 
    This formulation makes explicit the coupling between DQPSK symbol recovery and CSI estimation, and shows that a tracked CSI can be used as a channel prior to improve subsequent symbol decisions, which in turn produces a more reliable CSI sequence for radar processing. 

    \item We develop a posterior-probability-aware CSI tracker for DAB+ differential OFDM symbols. 
    The estimator uses the predicted CSI to cast DQPSK transition recovery as a maximum a posteriori (MAP) detection problem, converts the resulting posterior transition probability into an effective observation uncertainty, and fuses the channel prior with the instantaneous CSI observation through an optimal weighting derived in the linear minimum mean squared error (LMMSE) sense. 

    \item We also introduce a reliability-informed fusion strategy to enhance CSI estimation for for radar sensing.
    Unlike the fused CSI obtained above, which is aimed to minimize recursive symbol estimation errors, the CSI fusion here is designed to retain reliable dynamic channel innovations associated with much weaker targets. 

    \item {We analytically characterize the performance of the proposed framework. 
    The analysis provide theoretical guarantees of the performance improvement by the proposed design. Compared with open loop processing, the prediction based DQPSK transition estimation has a 3-dB SNR gain asymptotically; and the fused CSI estimation always has a smaller mean squared error (MSE), with CSI error variance caused by transition estimation error scaled by the squared fusion gain. The fusion gain is small, and close to zero.  
    }

\end{itemize}

We validate the framework using both DAB+ Mode-I simulations and experiments. 
At 5~dB SNR, it reduces the symbol error rate by about $63\%$ and improves tracking CSI normalized MSE by $15.1$~dB over the open-loop estimation in \cite{DAB_ofdm_sensing}. 
In random target scenes under different frequency-selective fading, it improves the median target-to-background ratio by more than $11$~dB at 4~dB SNR. 
In a representative three-target range-Doppler map, the minimum target-to-background ratio increases from $17.8$~dB for open-loop processing to $33.1$~dB with the proposed sensing CSI, corresponding to a gain of more than $15$~dB. Experiments conducted in Sydney, Australia also demonstrate range-Doppler map improvements achieved by the proposed CSI estimation scheme. 

The remainder of the paper is organized as follows. Section~\ref{sec:model} presents the DAB+ signal model and sensing-oriented problem formulation. 
Section~\ref{sec:proposed_receiver} develops the posterior-aware differential CSI tracker and sensing-output formation. Section~\ref{sec:analysis} analyzes the transition reliability, frequency-correlated prediction, posterior CSI fusion, and sensing-output behavior. Section~\ref{sec:simulation_results} presents simulation and experimental results. Section~\ref{sec:conclusion} concludes the paper.

\section{Signal Model and Problem Statement}
\label{sec:model}

We consider DAB+ Mode-I reception following the ETSI DAB physical-layer specifications \cite{ETSIEN3004012017}. The sampling rate is $f_s=2.048$~MHz, the FFT size is $N_{\mathrm{fft}}=2048$, the cyclic-prefix length is $N_{\mathrm{cp}}=504$, and the $K=1536$ active subcarriers are indexed by
\begin{equation} \label{eq:K}
\mathcal K=\{-768,\ldots,-1,1,\ldots,768\}.
\end{equation}
A frame contains one null symbol, with $N_{\rm null}=2656$ samples, followed by $M=76$ useful OFDM symbols, with the first useful symbol being the phase-reference symbol (PRS).

After frame synchronization, clock offset compensation, cyclic-prefix removal, FFT, and active subcarrier extraction, the received symbol on subcarrier $k$, useful-symbol index $m$, and frame index $f$ is modeled as
\begin{equation}
Y_{k,m,f}=H_{k,m,f}X_{k,m,f}+W_{k,m,f},
\label{eq:obs_model}
\end{equation}
where $X_{k,m,f}$ is the transmitted DAB+ symbol, $W_{k,m,f}\sim\mathcal{CN}(0,\sigma_0^2)$ denotes additive white Gaussian noise (AWGN) plus other residual disturbance, and $H_{k,m,f}$ is the CSI on the active OFDM grid. 
Considering $P$ propagation paths, the CSI can be written as
\begin{equation}
H_{k,m,f}=\sum_{p=0}^{P-1}a_p[m,f]
 e^{-\jmathi2\pi k\Delta f\tau_p[m,f]},
\label{eq:channel_expansion}
\end{equation}
where $\Delta f=1$~kHz is the subcarrier spacing, and $a_p[m,f]$ and $\tau_p[m,f]$ are the complex gain and delay of the $p$th path. Direct-path leakage, static clutter, and moving-target reflections are all included in \eqref{eq:channel_expansion}. Thus, $H_{k,m,f}$ can be estimated to infer target information.

Differential \dqpsk is used in \dab system. 
The PRS provides a known frame-level anchor $X_{k,0,f}=X_k^{\mathrm{PRS}}$,
whereas subsequent useful symbols obey the following differential encoding,
\begin{equation}
X_{k,m,f}=X_{k,m-1,f}D_{k,m,f},~ m\geq1,
\label{eq:dqpsk_model}
\end{equation}
with $D_{k,m,f}\in\mathcal Q$ and
\begin{equation}
\mathcal Q=\left\{e^{\jmathi(\pi/4+q\pi/2)}:q=0,1,2,3\right\}.
\label{eq:dqpsk_alphabet}
\end{equation}
Thus, to estimate $H_{k,m,f}$ for passive sensing, one must also recover \dab communication symbols $\{X_{k,m,f}\}$ from the observation in \eqref{eq:obs_model}.

A conventional open-loop approach detects the differential transition
and then reconstructs the absolute symbols as
\begin{equation}
\hat X_{k,m,f}=X_k^{\mathrm{PRS}}\prod_{u=1}^{m}\hat D_{k,u,f}, 
\label{eq:naive_symbol_chain}
\end{equation}
using the recursive relationship in \eqref{eq:dqpsk_model}, where $\hat D_{k,u,f}$ is generally estimated based on hard decision. Specifically, with $\hat X_{k,m-1,f}$ estimated from the previous symbol, we have $Y_{k,m-1,f}^*Y_{k,m,f}\approx|H_{k,m-1,f}|^2\hat X_{k,m-1,f}^*D_{k,m,f}$, where noise terms are suppressed for brevity, and hence $D_{k,m,f}$ is estimated by mapping $Y_{k,m-1,f}^*Y_{k,m,f}\big/\big(|H_{k,m-1,f}|^2\hat X_{k,m-1,f}^*\big)$ to the closest constellation point in $\mathcal Q$. 
If a transition slip occurs, the accumulated multiplicative symbol error becomes
\begin{equation}
\varepsilon_{k,m,f}=\frac{\hat X_{k,m,f}}{X_{k,m,f}}
=\prod_{u=1}^{m}\frac{\hat D_{k,u,f}}{D_{k,u,f}},
\label{eq:cumulative_error}
\end{equation}
and direct symbol removal gives
\begin{equation}
\hat H_{k,m,f}^{\mathrm{naive}}
=\frac{Y_{k,m,f}}{\hat X_{k,m,f}}
=H_{k,m,f}\varepsilon_{k,m,f}^{-1}
+\frac{W_{k,m,f}}{\hat X_{k,m,f}}.
\label{eq:naive_csi}
\end{equation}
Equation~\eqref{eq:naive_csi} 
explains how open-loop demodulation could fail passive sensing based on \dab CSI. 
From it, we see that a local DQPSK error is not a single-symbol outlier; it persists as a multiplicative CSI error until the next PRS anchor. Due to channel fading, such errors can be severe, making open-loop CSI estimation ineffective for passive radar sensing.

\section{Proposed Posterior-Probability-Aware Data-Aided CSI Estimation}
\label{sec:proposed_receiver}
\label{sec:method}

This section develops a robust CSI estimation scheme, converting the active subcarrier observations\footnote{As mentioned therein, we assume that the baseband samples captured by a passive radar receiver are preprocessed, including typically PRS-aided frame acquisition, sampling frequency offset correction, and CP-based carrier frequency offset
removal, to recover signals on active subcarriers of OFDM waveforms. One can find common preprocessing methods documented in related works \cite{DAB_ofdm_sensing,KaiWu2022Tutorial,kai_liu2024practical}. Here, we mainly focus on CSI estimation from the recovered subcarrier signals, as modeled in \eqref{eq:obs_model}.} $Y_{k,m,f}$ given in (\ref{eq:obs_model})
into a reliable CSI sequence for passive radar sensing.  
The key idea is to use the previous posterior CSI estimate as a one-step prior for DQPSK transition detection, and then to fuse the resulting CSI observation with channel prior efficiently.

\subsection{Frequency-Correlated CSI Prediction}
\label{subsec:channel_prediction}

For each frame or processing block indexed by $f$, define the CSI vector over the active DAB+ subcarriers as
\begin{equation}
    \bm H_{m,f}=\big[H_{k,m,f}\big]_{k\in\mathcal K}\in\mathbb C^{|\mathcal K|}.
    \label{eq:CSI_vector_def_proposed}
\end{equation}
The CSI is composed of the direct path, static multipath, slowly varying clutter, and weak moving-target echoes. Over adjacent useful OFDM symbols, the dominant components are usually locally predictable, while target-induced perturbations and residual impairments can be modeled as process uncertainty. In addition, because the CSI is the Fourier transform of a finite-delay channel given in \eqref{eq:channel_expansion}, neighboring subcarriers are statistically correlated rather than independent. These two facts motivate the following low-complexity prediction model,
\begin{equation}
    \bm H_{m,f}
    =
    \mathcal S_{\alpha}\!\left(\bm H_{m-1,f}\right)
    +\bm V_{m,f},
    ~
    \bm V_{m,f}\sim\mathcal{CN}\!\left(\bm 0,\bm Q_m\right),
    \label{eq:channel_prior_principled_enhanced}
\end{equation}
where $\mathcal S_{\alpha}(\cdot)$ is a smoothing operator across frequency and $\bm V_{m,f}$ captures process uncertainty due to moving scatterers, residual synchronization errors, front-end imperfections, thermal noise propagation, and model mismatch. A nearest-neighbor implementation adopted in this work is
\begin{equation}
    \left[\mathcal S_{\alpha}(\bm h)\right]_k
    =
    (1-\alpha)h_k
    +
    \frac{\alpha}{|\mathcal N_k|}
    \sum_{\ell\in\mathcal N_k}h_{\ell},
    ~ 0\leq \alpha<1,
    \label{eq:smoothing_operator_principled_enhanced}
\end{equation}
where $h$ denotes element in the input invetor $\bm h$, $\mathcal N_k$ is the set of adjacent active subcarriers of tone $k$.

Given the posterior estimate $\hat{\bm H}_{m-1,f}$ from the previous useful symbol, the predicted CSI used for data detection is
\begin{equation}
    \tilde{\bm H}_{m,f}
    =
    \mathcal S_{\alpha}\!\left(\hat{\bm H}_{m-1,f}\right),
    \label{eq:predicted_CSI_principle_enhanced}
\end{equation}
with $\tilde H_{k,m,f}$ denoting its $k$th entry.
Channel variations from, e.g., clock offset residuals and Doppler etc., are inevitable. These will be treated as prediction noise/uncertainty and suppressed by the proposed CSI fusion, as will be clear shortly. The frequency smoothing as enforced by $\mathcal S_{\alpha}(\cdot)$ helps improve CSI prediction quality. This is analyzed below. 

\begin{remark}
\label{lem:mild_frequency_coupling}
Consider an interior active subcarrier $k$ with
$\mathcal N_k=\{k-1,k+1\}$ in
\eqref{eq:smoothing_operator_principled_enhanced}. Suppose the
previous posterior CSI errors on neighboring tones are
zero-mean, locally uncorrelated, and have common variance
$\sigma_E^2$. Define the local CSI curvature as
\begin{equation}
    C_{k,m-1,f}
    =
    H_{k-1,m-1,f}-2H_{k,m-1,f}+H_{k+1,m-1,f}.
\end{equation}
Then, relative to the unsmoothed one-step predictor
$\hat H_{k,m-1,f}$, the smoothed predictor
$\tilde H_{k,m,f}$ in
\eqref{eq:predicted_CSI_principle_enhanced} reduces the
subcarrier-wise prediction MSE by
\begin{equation}
    \Delta_{\rm pred}(\alpha)
    =
    \left(2\alpha-\frac{3}{2}\alpha^2\right)\sigma_E^2
    -
    \frac{\alpha^2}{4}|C_{k,m-1,f}|^2 .
    \label{eq:prediction_mse_gain_alpha}
\end{equation}
See Appendix~\ref{app:frequency_prediction} for proof. 
Consequently, for sufficiently small $\alpha>0$, the predictor
always reduces MSE if the previous estimation
error dominates the local CSI curvature $C_{k,m-1,f}$. 

\end{remark}

\subsection{Prediction-Aided MAP Transition Detection}
\label{subsec:data_aided_observation}

Let $q_{k,m,f}$ denote the DQPSK transition from symbol $m-1$ to symbol $m$ on subcarrier $k$, i.e.,
\begin{equation}
    X_{k,m,f}=X_{k,m-1,f}q_{k,m,f},
    ~ q_{k,m,f}\in\mathcal Q,
    \label{eq:dqpsk_transition_model_enhanced}
\end{equation}
where $\mathcal Q$ is the transition alphabet in \eqref{eq:dqpsk_alphabet}.  For a candidate transition $q\in\mathcal Q$, the predicted CSI and previous reconstructed symbol imply the following likelihood
\begin{equation}
    Y_{k,m,f}\mid q
    \sim
    \mathcal{CN}\!\left(
    \tilde H_{k,m,f}\hat X_{k,m-1,f}q,
    \sigma_{\me,k,m,f}^{2}
    \right),
    \label{eq:transition_likelihood_principled_enhanced}
\end{equation}
where $\sigma_{\me,k,m,f}^{2}$ includes thermal noise and prediction uncertainty.  Define the residual as
\begin{equation}
    r_{k,m,f}(q)
    =
    \left|
    Y_{k,m,f}-\tilde H_{k,m,f}\hat X_{k,m-1,f}q
    \right|^2 .
    \label{eq:transition_residual_def}
\end{equation}

With proof given in Appendix~\ref{app:map_detector}, the following efficient optimal detector can be established for detecting the DQPSK transition.

\begin{proposition}
   \it  With equal transition priors and a common residual variance across the four hypotheses, the following minimization leads to an MAP detector for $q_{k,m,f}$
\begin{equation}
    \hat q_{k,m,f}
    =
    \arg\min_{q\in\mathcal Q} r_{k,m,f}(q).
    \label{eq:map_transition_principled_enhanced}
\end{equation}
\end{proposition}

The reconstructed symbol and the  CSI observation are then
\begin{equation}
    \hat X_{k,m,f}=\hat X_{k,m-1,f}\hat q_{k,m,f},
    \label{eq:symbol_update_principled_enhanced}
\end{equation}
\begin{equation}
    Z_{k,m,f}={Y_{k,m,f}}\big/{\hat X_{k,m,f}} .\label{eq:instantaneous_CSI_observation_enhanced}
\end{equation}
Since symbols over active DAB+ subcarriers have constant modulus, the division in \eqref{eq:instantaneous_CSI_observation_enhanced} does not change the additive-noise variance whether the estimated symbol is correct or not.

\subsection{Posterior-Probability-Aware CSI Update}
\label{subsec:posterior_probability_fusion}

The hard transition decision in \eqref{eq:map_transition_principled_enhanced} can make the CSI estimate suffer from ambiguous transition decision. 
Since $\tilde H_{k,m,f}$ is the prediction and $Z_{k,m,f}$ is based on the measurement $Y_{k,m,f}$, which is equivalent to an observation, we can apply a Kalman-like fusion to further enhance channel estimation.  
To do this, we start with analyzing the error of the two estimates. 

The prediction uncertainty is the difference between the prediction, $\tilde H_{k,m,f}$, and the true channel. Since the true channel is unknown, a reasonable approximation is the most recent stable estimation, which is the 
posterior CSI estimate $\hat H_{k,m-1,f}$ from the last symbol. Hence, the prediction uncertainty can be estimated as
\begin{equation}
    \sigma_{\sfp,k,m,f}^{2}
    \approx
    \left|\hat H_{k,m-1,f}-\tilde H_{k,m,f}\right|^{2}.
    \label{eq:prediction_variance_estimator}
\end{equation}
A local average of \eqref{eq:prediction_variance_estimator} over neighboring tones, similar to \eqref{eq:smoothing_operator_principled_enhanced}, can be applied to reduce isolated variance spikes. 

To analyze the observation uncertainty, let's start from the DQPSK likelihood in \eqref{eq:transition_likelihood_principled_enhanced}. The noise in $Y_{k,m,f}$ consists of prediction noise and white noise $W_{k,m,f}$; see \eqref{eq:obs_model}, where the variance of $W_{k,m,f}$ is $\sigma_0^2$. Thus, 
The overall noise variance in the DQPSK likelihood \eqref{eq:transition_likelihood_principled_enhanced} is
\begin{equation}
    \sigma_{\me,k,m,f}^{2}
    \approx
    \sigma_0^{2}+\sigma_{\sfp,k,m,f}^{2}.
    \label{eq:transition_residual_variance}
\end{equation}
Based on \eqref{eq:transition_likelihood_principled_enhanced}, the posterior probability of each candidate transition becomes
\begin{align}
    \pi_{k,m,f}(q)
    &\triangleq
    \Pr(q_{k,m,f}=q\mid Y_{k,m,f})
    \nonumber\\
    & =
    \frac{
    \exp\!\left(-r_{k,m,f}(q)/\sigma_{\me,k,m,f}^{2}\right)
    }{
    \sum_{q'\in\mathcal Q}
    \exp\!\left(-r_{k,m,f}(q')/\sigma_{\me,k,m,f}^{2}\right)
    } .
    \label{eq:dqpsk_posterior_probability_final}
\end{align}
As proved in Appendix~\ref{app:observation_variance}, the observation error of $Z_{k,m,f}$
can be quantified as follows.

\begin{proposition}
    \it 
    The observation variance of $Z_{k,m,f}$ is
\begin{equation}
    \sigma_{\mz,k,m,f}^{2}
    =
    \sigma_0^{2}
    +
    |\tilde H_{k,m,f}|^2
    \sum_{q\in\mathcal Q}
    \pi_{k,m,f}(q)
    \left|\frac{q}{\hat q_{k,m,f}}-1\right|^2 .
    \label{eq:posterior_observation_variance_final}
\end{equation}
\end{proposition}

With both prediction and observation errors analyzed above, we provide the following LMMSE fusion of the two CSI estimates. See Appendix~\ref{app:lmmse_fusion} for the proof.

\begin{proposition} \label{pp: lmmse}
    \it The LMMSE fusion of the predicted CSI and CSI observation is given by 
    \begin{equation}
    \hat H_{k,m,f}
    =
    \tilde H_{k,m,f}
    +K_{k,m,f}\left(Z_{k,m,f}-\tilde H_{k,m,f}\right),
    \label{eq:posterior_kalman_update}
\end{equation}
where the optimal fusion gain is 
\begin{equation}
    K_{k,m,f}
    =
    \frac{\sigma_{\sfp,k,m,f}^{2}}
    {\sigma_{\sfp,k,m,f}^{2}+\sigma_{\mz,k,m,f}^{2}}.
    \label{eq:posterior_kalman_gain}
\end{equation}
\end{proposition}

{The LMMSE update in Proposition~\ref{pp: lmmse} is optimal for subcarrier-wise CSI estimation under the conditional error model in Appendix~\ref{app:lmmse_fusion}. This estimate has the key role as a \emph{CSI tracker}. Specifically, it is fed back to \eqref{eq:predicted_CSI_principle_enhanced} as a prior and determines the quality of the next DQPSK transition test.} For this role, ``conservativeness'' is desirable. Namely, if the  observation $Z_{k,m,f}$ is unreliable because of receiver noise or DQPSK ambiguity, the fusion gain $K_{k,m,f}$ will be small to suppress the innovation $\left(Z_{k,m,f}-\tilde H_{k,m,f}\right)$. This, however, will also suppress sensing information, which is often weak. Therefore, we introduce another fusion below to help preserve sensing information.

\subsection{Reliability-Aware CSI Fusion to Enhance Radar Sensing}
\label{subsec:sensing_CSI}

Consider a general innovation gain $G$ and define
\begin{equation}
    \hat H_{k,m,f}(G)
    =
    \tilde H_{k,m,f}
    +G\left(Z_{k,m,f}-\tilde H_{k,m,f}\right).
    \label{eq:generic_gain_CSI_output}
\end{equation}
The ordinary tracking MSE part of the fusion is
$(1-G)^2\sigma_{\sfp,k,m,f}^{2}
    +G^2\sigma_{\mz,k,m,f}^{2}$.
To help preserve sensing information, we add a penalty for attenuating dynamic innovation that is likely to be reliable for sensing,
\begin{equation}
\begin{aligned}
    J_{\rm sen}(G)
    ={}&(1-G)^2\sigma_{\sfp,k,m,f}^{2}
    +G^2\sigma_{\mz,k,m,f}^{2}
    +\chi_{k,m,f}(1-G)^2,
\end{aligned}\nonumber
\end{equation}
where $\chi_{k,m,f}\geq0$ is an equivalent sensing-innovation retention weight. 
The last term penalizes excessive suppression of the innovation that often contain radar target information. Taking the derivative of $J_{\rm sen}(G)$ with respect to $G$ and equating the result to zero give
\begin{equation}
    G_{k,m,f}
    =
    \frac{
    \sigma_{\sfp,k,m,f}^{2}+\chi_{k,m,f}
    }{
    \sigma_{\sfp,k,m,f}^{2}+\sigma_{\mz,k,m,f}^{2}+\chi_{k,m,f}
    } .
    \label{eq:sensing_gain_regularised_form}
\end{equation}
When $\chi_{k,m,f}=0$, the sensing gain reduces to the tracking gain in \eqref{eq:posterior_kalman_gain}; as $\chi_{k,m,f}$ increases, the solution moves toward direct innovation preservation.

Through some basic math manipulation, the form in \eqref{eq:sensing_gain_regularised_form} can be written as a relaxation from the tracking gain to the observation gain,
\begin{equation}
    G_{k,m,f}
    =
    K_{k,m,f}
    +\eta_{k,m,f}\left(1-K_{k,m,f}\right),
    \label{eq:sensing_gain_eta_form}
\end{equation}
where the coefficient $\eta_{k,m,f}$ is given by 
\begin{equation}
    \eta_{k,m,f}
    =
    \frac{\chi_{k,m,f}}
    {\sigma_{\sfp,k,m,f}^{2}+\sigma_{\mz,k,m,f}^{2}+\chi_{k,m,f}}
    \in[0,1].
    \label{eq:sensing_eta_definition}
\end{equation}
This form gives a useful interpretation. The tracking LMMSE gain
\(K_{k,m,f}\) admits only the MSE-optimal fraction of the innovation
\(Z_{k,m,f}-\tilde H_{k,m,f}\) into the recursive state, and rejects the
remaining fraction \(1-K_{k,m,f}\). The sensing gain recovers a
controlled portion \(\eta_{k,m,f}\) of this rejected innovation for
radar sensing. From \eqref{eq:sensing_gain_eta_form} and \eqref{eq:sensing_eta_definition}, \(G_{k,m,f}\) lies
between the conservative tracking gain and the direct-observation gain,
\[
K_{k,m,f}\leq G_{k,m,f}\leq 1.
\]
Therefore, the sensing output preserves reliable
target-induced slow-time variations, but does not amplify the
 observation beyond direct CSI division.

The remaining question is how to choose \(\eta_{k,m,f}\). 
For sensing, it is difficult to find a reliable approximation of the true state, as sensing information is much weaker than direct path and other clutters. Thus, we introduce a reliability based weight construction. 
Since the
innovation is useful for sensing only when the DQPSK transition decision
is reliable, we employ the posterior
concentration of the selected transition, as depicted by $\pi_{k,m,f}(\hat q_{k,m,f})$ in \eqref{eq:dqpsk_posterior_probability_final}, to assign \(\eta_{k,m,f}\) as 
\begin{equation}
    \eta_{k,m,f}
    =
    \left[
    \frac{
    \pi_{k,m,f}(\hat q_{k,m,f})-1/|\mathcal Q|
    }{
    1-1/|\mathcal Q|
    }
    \right]_{0}^{1}, \label{eq:sensing_reliability_score}
\end{equation}
where \([x]_{0}^{1}\triangleq\min\{1,\max\{0,x\}\}\).
When the posterior $\pi_{k,m,f}(\hat q_{k,m,f})$ is uninformative, we have $\pi_{k,m,f}(\hat q_{k,m,f})=1/|\mathcal Q|$. This gives \(\eta_{k,m,f}=0\), with the sensing
CSI reduced to the tracking CSI. When the selected
transition is highly reliable, \(\eta_{k,m,f}\) approaches one, and the
sensing CSI emphasizes direct observation, preserving
target-bearing innovation.

\begin{algorithm}[!t]
\caption{\small Posterior-Aware CSI Tracking and Estimation}
\label{alg:ppaware_CSI_algorithm}
\centering
{\footnotesize
\begin{minipage}{0.94\linewidth}
\noindent{Input:} active subcarrier observations $Y_{k,m,f}$, PRS symbols $X^{\rm PRS}_{k}$, smoothing factor $\alpha$, post-FFT noise variance $\sigma_0^2$.

\vspace{0.5mm}
\noindent{Output:} Reconstructed data symbols $\hat X_{k,m,f}$, recursive tracking CSI $\hat{\bm H}^{\rm tr}_{m,f}$, sensing-output CSI $\hat{\bm H}^{\rm sen}_{m,f}$

\vspace{0.5mm}
\begin{enumerate}[leftmargin=*]
    \item {PRS initialization:} For each frame $f$, set $\hat X_{k,0,f}=X^{\rm PRS}_{k},~
        \hat H^{\rm tr}_{k,0,f}
        =
        \hat H^{\rm sen}_{k,0,f}
        =
        \frac{Y_{k,0,f}}{X^{\rm PRS}_{k}}$.

    \item {For each useful OFDM symbol $m\geq 1$:}
    \begin{enumerate}[leftmargin=*]
        \item {CSI prediction:} Form the one-step prediction $\tilde{\bm H}_{m,f}
            =
            \mathcal S_{\alpha}\!\left(
            \hat{\bm H}^{\rm tr}_{m-1,f}
            \right)$
        following \eqref{eq:predicted_CSI_principle_enhanced}. 

        \item {Prediction uncertainty:} Estimate the prediction error
        $\sigma_{\sfp,k,m,f}^{2}$ per
        \eqref{eq:prediction_variance_estimator}, and set the DQPSK likelihood error
        $\sigma_{\me,k,m,f}^{2}$ by
        \eqref{eq:transition_residual_variance}.

        \item {Prediction-aided transition detection:} For each
        $q\in\mathcal Q$, compute $r_{k,m,f}(q)$ using
        \eqref{eq:transition_residual_def}, and select
        $\hat q_{k,m,f}$ using the MAP rule in
        \eqref{eq:map_transition_principled_enhanced}.

        \item {Symbol reconstruction and  CSI observation:}
        Update
        $\hat X_{k,m,f}$ using
        \eqref{eq:symbol_update_principled_enhanced}, and form
        $Z_{k,m,f}$ using
        \eqref{eq:instantaneous_CSI_observation_enhanced}.

        \item {Posterior transition reliability:} Compute
        $\pi_{k,m,f}(q)$ from
        \eqref{eq:dqpsk_posterior_probability_final}, and obtain the effective observation variance
        $\sigma_{\mz,k,m,f}^{2}$ from
        \eqref{eq:posterior_observation_variance_final}.

        \item {Recursive tracking update:} Compute the posterior LMMSE gain
        $K_{k,m,f}$ from \eqref{eq:posterior_kalman_gain}, and update the recursive tracking CSI
        \[
            \hat H^{\rm tr}_{k,m,f}
            =
            \tilde H_{k,m,f}
            +
            K_{k,m,f}
            \left(
            Z_{k,m,f}
            -
            \tilde H_{k,m,f}
            \right).
        \]
        This estimate is fed back to the next-symbol prediction.

        \item {Sensing-output formation:} Compute the posterior reliability coefficient $\rho_{k,m,f}$ and the sensing gain $G_{k,m,f}$ according to \eqref{eq:sensing_gain_eta_form} and
        \eqref{eq:sensing_reliability_score}, and form
        \[
            \hat H^{\rm sen}_{k,m,f}
            =
            \tilde H_{k,m,f}
            +
            G_{k,m,f}
            \left(
            Z_{k,m,f}
            -
            \tilde H_{k,m,f}
            \right).
        \]
    \end{enumerate}

\end{enumerate}
\end{minipage}
}
\end{algorithm}

\subsection{Overall CSI Estimation Procedure}

Algorithm~\ref{alg:ppaware_CSI_algorithm} summarizes the proposed
posterior-probability-aware CSI tracking and estimation scheme. The algorithm has two
distinct CSI outputs. The first is the tracking CSI
$\hat{\bm H}^{\rm tr}_{m,f}$, as derived in Section \ref{subsec:posterior_probability_fusion} and obtained by Step 2f). It is the posterior LMMSE state used by
the next-symbol predictor; Step 2a). 
The second output is the sensing CSI
$\hat{\bm H}^{\rm sen}_{m,f}$, as derived in Section \ref{subsec:sensing_CSI} and obtained in Step 2g). 
It is generated from the same
prediction, observation, and posterior transition
statistics, as calculated in Steps 2b)-2e). 
We add superscripts in the algorithm to differentiate the two CSI outputs. The input $\sigma_0^2$ denotes the post-FFT complex noise variance per
active resource element. It can be estimated from
inactive subcarriers, guard bands, silent/null intervals, or high-delay CIR bins outside the expected channel-delay
support.

\subsection{Range-Doppler Map}
To facilitate performance evaluation, we provide minimum operation to generate a range-Doppler map from the sensing CSI $H^{{\rm sen}}_{\kappa,m,f}$ produced by Algorithm \ref{alg:ppaware_CSI_algorithm}.
Advanced algorithms, such as correlation matrix based high/super resolution methods \cite{book_van2004optimum}, can be applied, which is beyond the scope of this work. 

For each $(m,f)$, the active subcarrier estimate is first embedded into the full FFT grid as
\begin{equation}
    \bar H^{{\rm sen}}_{\kappa,m,f}
    =
    \begin{cases}
    \hat H^{{\rm sen}}_{k,m,f}, & \kappa=k\%N_{\rm fft},~k\in\mathcal K,\\
    0, & \text{otherwise},
    \end{cases}
    \label{eq:sensing_CSI_full_grid}
\end{equation}
where $\mathcal K$ is given in \eqref{eq:K}, and $k\%N_{\rm fft}$ denotes modulo operation, mapping the signed active subcarrier index to the FFT-bin index\footnote{Similar to fftshift in matlab or python programming.}. The channel impulse response (CIR) is obtained by
\begin{equation}
    h^{{\rm sen}}_{r,m,f}
    =
    \left[\bm F_{N_{\rm fft}}^{\rm H}\bar{\bm H}^{{\rm sen}}_{m,f}\right]_{r},
    ~ 0\leq r<N_{\rm fft},
    \label{eq:sensing_cir_from_CSI}
\end{equation}
where $\bm F_{N_{\rm fft}}$ denotes the unitary DFT matrix. Since the direct path and static clutter dominate many delay bins, a slow-time difference is taken before Doppler processing,
\begin{equation}
    \breve h^{{\rm sen}}_{r,m,f}
    =
    h^{{\rm sen}}_{r,m,f}-h^{{\rm sen}}_{r,m-1,f}, ~\forall f.
    \label{eq:sensing_cir_background_removal}
\end{equation}

Let $t_{m,f}=fT_{\rm fr}+mT_{\rm sym}$ denote the DAB slow-time grid and let $w_{m,f}$ be an optional slow-time taper, where $T_{\rm sym}=(N_{\rm fft}+N_{\rm cp})/f_s$ and $T_{\rm fr}=MT_{\rm sym}+N_{\rm null}/f_s$. For each delay bin $r$ in the useful range set $\mathcal R$ determined by the cyclic prefix support, the range-Doppler map is formed as
\begin{equation}
    \mathcal G(r,\nu)
    =
    \sum_{f=0}^{F-1}\sum_{m=0}^{M-2}
    w_{m,f}\breve h^{{\rm sen}}_{r,m,f}
    e^{-j2\pi\nu t_{m,f}} ,
    \label{eq:sensing_range_doppler_map}
\end{equation}
where the range of $m$ is reduced by one due to the difference in \eqref{eq:sensing_cir_background_removal}. 
The delay index corresponds to bistatic range $R_r=cr/f_s$, while the Doppler coordinate $\nu$ can be converted to a radial or bistatic velocity according to the relevant geometry, e.g., $v=\lambda\nu/\gamma_{\rm B}$ under the bistatic scale factor $\gamma_{\rm B}$. Depending on geometry, $\gamma_{\rm B}\in [1,2]$.

\section{Performance Analysis}
\label{sec:analysis}

This section analyzes the performance gains produced by the receiver in Section~\ref{sec:proposed_receiver}. We quantify improvements of DQPSK transition estimation and CSI estimation.

\subsection{Transition SNR and DQPSK Slip Probability}
\label{subsec:transition_snr_analysis}

For a general DQPSK transition detector with conditional observation
\begin{equation}
 \zeta=a q+u,
 ~ q\in\mathcal Q,
 ~ u\sim\mathcal{CN}(0,\sigma_u^2),
 \label{eq:generic_dqpsk_observation}
\end{equation}
the four hypotheses in $\mathcal Q$ have minimum Euclidean separation $\sqrt{2}|a|$. The pairwise error probability is upper-bounded by $Q(|a|/\sigma_u)$, and the union bound gives \cite{book_mimo_ofdm}
\begin{equation}
 P_{\rm slip}
 \leq
 3Q\!\left(\sqrt{\Gamma}\right),
 ~
 \Gamma\triangleq\frac{|a|^2}{\sigma_u^2}.
 \label{eq:generic_slip_bound}
\end{equation}
The relevant quantity for DAB+ sensing is not only the received SNR, but the transition SNR $\Gamma$ after channel uncertainty and receiver processing are included.

For the proposed detector in \eqref{eq:transition_likelihood_principled_enhanced}, $a=\tilde H_{k,m,f}\hat X_{k,m-1,f}$ and $\sigma_u^2=\sigma_{\me,k,m,f}^2$. Since $|\hat X_{k,m-1,f}|=1$, the effective transition SNR is
\begin{equation}
 \Gamma^{\rm prop}_{k,m,f}
 =
 \frac{|\tilde H_{k,m,f}|^2}
 {\sigma_0^2+\sigma_{\sfp,k,m,f}^{2}}.
 \label{eq:prop_effective_transition_snr}
\end{equation}
Thus the proposed slip probability satisfies
\begin{equation}
 P^{\rm prop}_{\rm slip}(k,m,f)
 \leq
 3Q\!\left(
 \sqrt{\frac{|\tilde H_{k,m,f}|^2}{\sigma_0^2+\sigma_{\sfp,k,m,f}^{2}}}
 \right).
 \label{eq:prop_slip_bound}
\end{equation}
Equation~\eqref{eq:prop_slip_bound} directly exposes the two performance levers of the proposed method. Specifically, a reliable predicted CSI increases the DQPSK decision distance, while a smaller prediction uncertainty reduces the effective disturbance variance.

It is useful to compare \eqref{eq:prop_effective_transition_snr} with the conventional open-loop differential statistic
\begin{equation}
 R^{\rm ol}_{k,m,f}=Y_{k,m,f}Y^{*}_{k,m-1,f}.
 \label{eq:open_loop_differential_statistic}
\end{equation}
When $H_{k,m,f}\approx H_{k,m-1,f}\triangleq H_{k,m,f}^{\rm loc}$, where $H_{k,m,f}^{\rm loc}$ denotes a locally stable channel, and DAB symbols have unit modulus,
\begin{equation}
 R^{\rm ol}_{k,m,f}
 =
 |H_{k,m,f}^{\rm loc}|^2 q_{k,m,f}+\xi_{k,m,f},
 \label{eq:open_loop_differential_model}
\end{equation}
where, to the first order,
\begin{equation}
 \mathbb E\{|\xi_{k,m,f}|^2\}
 \approx
 2|H_{k,m,f}^{\rm loc}|^2\sigma_0^2+\sigma_0^4+\sigma_{\Delta H,k,m,f}^{2}.
 \label{eq:open_loop_noise_variance}
\end{equation}
Here $\sigma_{\Delta H,k,m,f}^{2}$ collects the modeling error due to the difference between $H_{k,m,f}H_{k,m-1,f}^{*}$ and the locally constant approximation. The corresponding transition SNR is
\begin{equation}
 \Gamma^{\rm ol}_{k,m,f}
 \approx
 \frac{|H_{k,m,f}^{\rm loc}|^4}
 {2|H_{k,m,f}^{\rm loc}|^2\sigma_0^2+\sigma_0^4+\sigma_{\Delta H,k,m,f}^{2}}.
 \label{eq:open_loop_effective_transition_snr}
\end{equation}
At high SNR and under a locally static channel, \eqref{eq:open_loop_effective_transition_snr} approaches $|H|^2/(2\sigma_0^2)$, whereas \eqref{eq:prop_effective_transition_snr} approaches $|H|^2/\sigma_0^2$ when the prediction error is negligible. Thus prediction-aided detection gives an inherent $3$~dB transition SNR advantage over the open-loop differential statistic based on two noisy symbols, before considering the additional gain from frequency-domain prediction and posterior fusion.

\subsection{Gain from Frequency-Correlated Prediction}
\label{subsec:prediction_gain_analysis}

Let $P_{m-1}$ denote the variance of the previous subcarrier-wise CSI estimation error, $Q_m$ the temporal process-innovation variance, and $B_{\alpha,k,m,f}$ the squared curvature bias induced by \eqref{eq:smoothing_operator_principled_enhanced}. From Appendix~\ref{app:frequency_prediction}, the smoothed prediction variance can be approximated as
\begin{equation}
 P^{-}_{\alpha,k,m,f}
 \approx
 \rho_{\alpha}P_{m-1}+Q_m+B_{\alpha,k,m,f},
 \label{eq:smoothed_prediction_variance}
\end{equation}
where $\rho_{\alpha}=(1-\alpha)^2+\frac{\alpha^2}{2}$.
Without frequency smoothing, the corresponding variance is approximately
\begin{equation}
 P^{-}_{0,k,m,f}
 \approx
 P_{m-1}+Q_m.
 \label{eq:unsmoothed_prediction_variance}
\end{equation}
Assuming the smoothing operation does not appreciably change the local CSI magnitude, the transition-SNR gain over the unsmoothed predictor is
\begin{equation}
 G_{\Gamma,\alpha}
 \triangleq
 \frac{\Gamma_{\alpha}}{\Gamma_{0}}
 \approx
 \frac{\sigma_0^2+P^{-}_{0,k,m,f}}
 {\sigma_0^2+P^{-}_{\alpha,k,m,f}}\approx
 \frac{\sigma_0^2+P_{m-1}+Q_m}
 {\sigma_0^2+\rho_{\alpha}P_{m-1}+Q_m}, \nonumber
\end{equation}
where the last approximation is due to small curvature bias.
For $\alpha=0.15$, $\rho_{\alpha}=0.734$, corresponding to up to $10\log_{10}(1/\rho_{\alpha})=1.34$~dB prediction-error-variance reduction in the estimation-error-dominated regime. The gain diminishes when thermal noise or temporal innovation dominates, and it reverses if $B_{\alpha,k,m,f}$ becomes large. This indicates that moderate smoothing improves transition detection in frequency-selective fading, while excessive smoothing degrades delay-domain sensing by biasing the CSI.

Combining \eqref{eq:generic_slip_bound} and $G_{\Gamma,\alpha}$,
the slip-rate improvement from smoothing satisfies approximately
\begin{equation}
 \frac{P^{(\alpha)}_{\rm slip}}{P^{(0)}_{\rm slip}}
 \lesssim
 \frac{Q\!\left(\sqrt{G_{\Gamma,\alpha}\Gamma_0}\right)}
 {Q\!\left(\sqrt{\Gamma_0}\right)},
 \label{eq:slip_reduction_from_alpha}
\end{equation}
where $\Gamma_0$ is the transition SNR without smoothing. Because the $Q$-function tail is exponential, even a modest transition-SNR gain can produce a much larger reduction in DQPSK slips in the medium-SNR regime.

\subsection{CSI MSE Reduction by Posterior-Aware Fusion}
\label{subsec:CSI_mse_fusion_analysis}

For a given subcarrier, symbol, and frame, denote the prediction-error variance by
\begin{equation}
 P^{-}_{k,m,f}\triangleq \sigma_{\sfp,k,m,f}^{2}
 \label{eq:prediction_variance_short}
\end{equation}
and the observation error variance by
\begin{equation}
 R_{k,m,f}\triangleq \sigma_{\mz,k,m,f}^{2}.
 \label{eq:observation_variance_short}
\end{equation}
Appendix~\ref{app:lmmse_fusion} gives the posterior CSI variance
\begin{equation}
 P^{+}_{k,m,f}
 =
 \frac{P^{-}_{k,m,f}R_{k,m,f}}{P^{-}_{k,m,f}+R_{k,m,f}}.
 \label{eq:posterior_CSI_mse_analysis}
\end{equation}
Relative to direct use of the observation $Z_{k,m,f}$, the MSE reduction factor is
\begin{equation}
 \frac{P^{+}_{k,m,f}}{R_{k,m,f}}
 =
 \frac{P^{-}_{k,m,f}}{P^{-}_{k,m,f}+R_{k,m,f}}
 =K_{k,m,f},
 \label{eq:mse_reduction_over_observation}
\end{equation}
where $K_{k,m,f}$ is the gain in \eqref{eq:posterior_kalman_gain}. Relative to relying only on the prediction, the MSE reduction factor is
\begin{equation}
 \frac{P^{+}_{k,m,f}}{P^{-}_{k,m,f}}
 =
 \frac{R_{k,m,f}}{P^{-}_{k,m,f}+R_{k,m,f}}
 =1-K_{k,m,f}.
 \label{eq:mse_reduction_over_prediction}
\end{equation}
Thus the update is never worse than either input under the conditional uncorrelated-error model, and its gain automatically selects the more reliable information source.

The contribution of DQPSK estimation error to the posterior CSI error is attenuated by $K_{k,m,f}^2$. Specifically, the ambiguity-induced part of the observation error in \eqref{eq:app_symbol_ambiguity_power} (in Appendix \ref{app:observation_variance}) contributes at most
\begin{equation}
 \varepsilon^{+}_{\rm amb}(k,m,f)
 \leq
 4K_{k,m,f}^{2}|\tilde H_{k,m,f}|^2\eta_{k,m,f},
 \label{eq:posterior_ambiguity_injection_bound}
\end{equation}
to the posterior CSI error power. Without posterior-aware fusion, the same term would enter with $K_{k,m,f}=1$. This is the principal CSI-domain performance advantage of the proposed receiver over MAP-only hard channel division, i.e., \eqref{eq:instantaneous_CSI_observation_enhanced}.

\section{Simulation and Experimental Results}
\label{sec:simulation_results}

This section validates the proposed communication-centric DAB+ sensing scheme using controlled Monte Carlo simulations and real life experiments. 

\begin{figure*}[!t]
\centering
\includegraphics[width=0.9\textwidth]{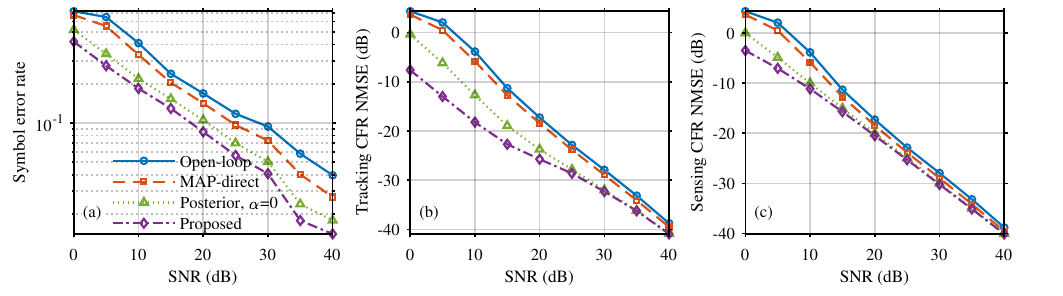}
\caption{
(a) reconstructed-symbol error rate, 
(b) tracking-CSI NMSE and 
(c) sensing-CSI NMSE. 
The proposed scheme uses $\alpha=0.15$, whereas the posterior baseline uses $\alpha=0$ to isolate the contribution of frequency-domain smoothing. The same legend applies to all.}
\label{fig:snr_scheme_tracking}
\end{figure*}

\begin{figure*}[!t]
\centering
\includegraphics[width=0.9\textwidth]{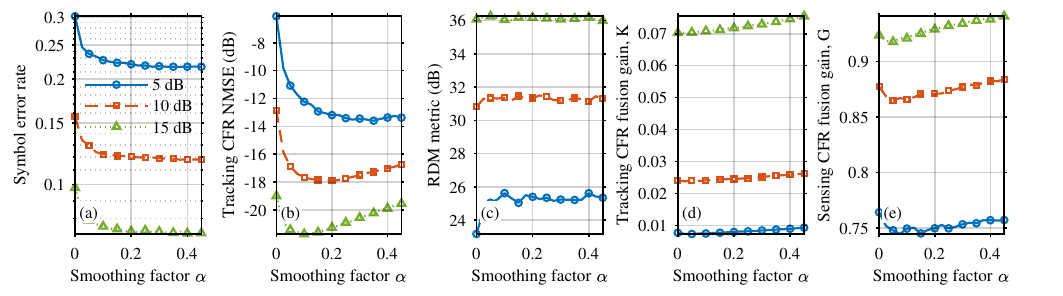}
\caption{Impact of the frequency-domain smoothing factor $\alpha$ on
(a) reconstructed-symbol error rate, (b) tracking-CSI NMSE, (c) TBR of the
resulting range-Doppler map, (d) average tracking fusion gain $\bar K$ and
(e) average sensing fusion gain $\bar G$. The same legend applies to all
subfigures.}
\label{fig:alpha_smoothing_tradeoff}
\end{figure*}

\subsection{Simulation Setup and Benchmark Schemes}
\label{subsec:simulation_setup}

The simulator follows the DAB+ Mode-I numerology in Section~\ref{sec:model}: $f_s=2.048$~MHz, $N_{\rm null}=2565$, $N_{\rm fft}=2048$, $N_{\rm cp}=504$, subcarrier spacing $\Delta f=1$~kHz and $M=76$ useful OFDM symbols per frame, with the first useful symbol serving as the PRS anchor. The post-PRS symbols are generated according to the differential rule in \eqref{eq:dqpsk_model}. The simulated observation follows \eqref{eq:obs_model}, where additive circular Gaussian noise is set by the desired post-FFT SNR. 
The propagation model contains a dominant direct component, static multipath components and one or more moving targets.

Four schemes are compared. 
\begin{enumerate}
    \item The \emph{open-loop} scheme uses the conventional two-symbol differential statistic in \eqref{eq:open_loop_differential_statistic}, followed by direct CSI division. This represents the prior art \cite{DAB_ofdm_sensing}.

    \item The \emph{MAP-direct} scheme uses the prediction-aided MAP transition rule in \eqref{eq:map_transition_principled_enhanced}, but directly admits the instantaneous CSI observation $Z_{k,m,f}$ in \eqref{eq:instantaneous_CSI_observation_enhanced}.
    \item The \emph{posterior, $\alpha=0$} scheme uses posterior-aware fusion but disables the frequency smoothing in \eqref{eq:smoothing_operator_principled_enhanced}.
    \item The \emph{proposed} scheme uses both posterior-aware fusion and frequency-smoothed prediction and forms sensing CSI, as illustrated in Algorithm \ref{alg:ppaware_CSI_algorithm}.
\end{enumerate}
Performance metrics include the subcarrier symbol error rate,
\begin{equation}
  P_{\rm sym}
  =
  \frac{1}{FK(M-1)}
  \sum_f\sum_{m=1}^{M-1}\sum_{k\in\mathcal K}
  \mathbf 1\{\hat X_{k,m,f}\neq X_{k,m,f}\},\nonumber
\end{equation}
and the tracking CSI normalized MSE (NMSE)
\begin{equation}
  {\rm NMSE}_{\rm CSI}
  =
  \frac{\sum_{f,m}\|\hat{\bm H}_{m,f}^{\rm tr}-\bm H_{m,f}\|_2^2}
  {\sum_{f,m}\|\bm H_{m,f}\|_2^2},\nonumber
\end{equation}
where $\hat{\bm H}_{m,f}^{\rm tr}$ is obtained in Step 2f) of Algorithm \ref{alg:ppaware_CSI_algorithm}.
The sensing-related metrics are computed from the range-Doppler map $\mathcal G(r,\ell)$ obtained in \eqref{eq:sensing_range_doppler_map}, where $r$ and $\ell$ denote the range- and Doppler-bin indices, respectively. Let $(r_0,\ell_0)$ be the nearest true target bin and let $\ell_{\rm dc}$ denote the zero-Doppler bin. We define the target mainlobe support, local evaluation support and zero-Doppler guard as
\begin{align}
\mathcal M
&=
\{(r,\ell): |r-r_0|\leq G_r,\ |\ell-\ell_0|\leq G_\ell\}, \nonumber\\
\mathcal W
&=
\{(r,\ell): |r-r_0|\leq W_r,\ |\ell-\ell_0|\leq W_\ell\}, \nonumber\\
\mathcal Z
&=
\{(r,\ell): |\ell-\ell_{\rm dc}|\leq G_{\rm dc}\}.\nonumber
\end{align}
The target-to-background ratio (TBR) is a key metric evaluating range-Doppler map (RDM) performance,
\begin{equation}
  {\rm TBR}
  =
  10\log_{10}
  \frac{
  \max_{(r,\ell)\in\mathcal M}|\mathcal G(r,\ell)|^2}
  {
  \frac{1}{|\mathcal B|}
  \sum_{(r,\ell)\in\mathcal B}|\mathcal G(r,\ell)|^2},
  \nonumber
\end{equation}
where $\mathcal B=\mathcal W\setminus(\mathcal M\cup\mathcal Z)$ is the local background region after excluding the target mainlobe and the residual zero-Doppler clutter guard.
To further distinguish range and Doppler smearing, we also evaluate target-anchored one-dimensional focus metrics. Define the local range cut and Doppler cut as
\begin{align}
  g_r(r)
  =
  \max_{\ell:\,|\ell-\ell_0|\leq G_\ell}
  |\mathcal G(r,\ell)|,
  ~g_\ell(\ell)
  =
  \max_{r:\,|r-r_0|\leq G_r}
  |\mathcal G(r,\ell)|.\nonumber
\end{align}
The range-focus metric is
\begin{equation}
  {\rm RF}
  =
  20\log_{10}
  \frac{
  \max_{(r,\ell)\in\mathcal M}|\mathcal G(r,\ell)|}
  {
  \sqrt{
  \frac{1}{|\mathcal R_{\rm s}|}
  \sum_{r\in\mathcal R_{\rm s}} g_r^2(r)}
  },
  \nonumber
\end{equation}
where
$\mathcal R_{\rm s}
=
\{r: |r-r_0|\leq W_r,\ |r-r_0|>G_r\}.$
Similarly, the Doppler-focus metric is
\begin{equation}
  {\rm DF}
  =
  20\log_{10}
  \frac{
  \max_{(r,\ell)\in\mathcal M}|\mathcal G(r,\ell)|}
  {
  \sqrt{
  \frac{1}{|\mathcal D_{\rm s}|}
  \sum_{\ell\in\mathcal D_{\rm s}} g_\ell^2(\ell)}
  },
  \nonumber
\end{equation}
where $\mathcal D_{\rm s}
=
\{\ell: |\ell-\ell_0|\leq W_\ell,\ |\ell-\ell_0|>G_\ell,\ |\ell-\ell_{\rm dc}|>G_{\rm dc}\}.$
RF measures whether target energy is concentrated in the correct delay cells, whereas DF measures whether it is concentrated in the correct Doppler cells. In the simulations, unless otherwise stated, we set $G_r=G_\ell=G_{\rm dc}=2$, $W_r=38$ and $W_\ell=26$. These are taken for relatively reliable estimation of the above statistics, hence robust sensing performance evaluation. In practice, one should carefully relate them with detecting performance pursued.

\subsection{Symbol and CSI Estimation Versus SNR}
\label{subsec:simulation_snr_results}

Fig.~\ref{fig:snr_scheme_tracking} shows the reconstructed-symbol error rate, tracking-CSI NMSE and sensing-CSI NMSE versus SNR. 
At 5~dB SNR (post FFT of $N_{\rm fft}=2048$ points), the proposed scheme reduces the symbol error rate from approximately $0.887$ for open-loop processing to $0.325$, corresponding to a $63\%$ reduction. At the same SNR, the tracking-CSI NMSE improves from about $2.1$~dB to $-13.0$~dB, i.e., a $15.1$~dB gain over open-loop processing and a $13.5$~dB gain over MAP-direct processing. 
At 10~dB SNR, the proposed scheme reduces the symbol error rate by about $61\%$ relative to open-loop processing and improves tracking-CSI NMSE by about $14.3$~dB. 
These performance gains benefit from the proposed design by fusing the frequency-smoothed prior and MAP-enabled CSI estimates.

\begin{figure*}[!t]
\centering
\includegraphics[width=0.9\textwidth]{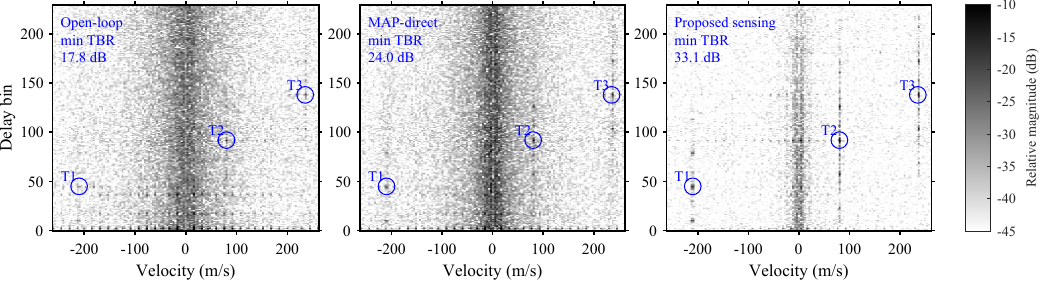}
\caption{Representative three-target range-Doppler maps at 5~dB SNR using the full active DAB+ subcarrier grid:
(a) open-loop differential processing,
(b) MAP-direct processing and
(c) the proposed sensing CSI.
The three target locations are marked by blue circles. 
}
\label{fig:rdm_maps_representative_scene}
\end{figure*}

The additional separation between the posterior $\alpha=0$ curve and the proposed curve further confirms the role of frequency-smoothed prediction in \eqref{eq:smoothing_operator_principled_enhanced}. Neighboring subcarriers provide a stabilizing prior in frequency-selective fading, reducing both transition estimation errors and error propagation along the DQPSK symbol chain.
The advantage is particularly significant at low and moderate SNRs, where a single open-loop or MAP-direct transition error can corrupt the subsequent symbol estimation until the next PRS re-anchor, as discussed after \eqref{eq:cumulative_error}. MAP-direct improves the transition decision relative to open-loop processing, but it directly admits the noisy observation $Z_{k,m,f}$ into the CSI estimate; hence, its tracking-CSI NMSE remains close to the open-loop curve. In contrast, the proposed scheme jointly improves the transition decision and the CSI estimation.

The sensing-CSI NMSE in Fig.~\ref{fig:snr_scheme_tracking}(c) follows the same trend but with a smaller absolute gain than the recursive tracking CSI. For example, at 5~dB SNR the proposed sensing output improves the CSI NMSE by about $9.3$~dB over open-loop processing, compared with $15.1$~dB for the tracking CSI. This behavior is expected. The tracking CSI is deliberately conservative because it is fed back into the next-symbol prediction, whereas the sensing CSI in \eqref{eq:sensing_gain_eta_form} re-admits reliable innovation to retain moving-target content. The advantage of sensing CSI over tracking CSI will be obvious shortly when discussing the radar RDM performance.

Fig.~\ref{fig:alpha_smoothing_tradeoff} studies the smoothing factor $\alpha$ in
\eqref{eq:smoothing_operator_principled_enhanced}. 
The results agree with the
transition SNR analysis in Section \ref{subsec:prediction_gain_analysis}. Moderate
frequency-domain smoothing reduces the local prediction noise variance and
improves transition reliability, whereas excessive smoothing provides little
additional benefit and may introduce curvature bias in frequency-selective
channels. The impact is most visible at low and medium SNR. 
At 5~dB SNR,
increasing $\alpha$ from zero to $0.1$ reduces the symbol error rate from
approximately $0.36$ to $0.258$ and improves tracking-CSI NMSE from about
$-6.1$~dB to $-12.2$~dB. At 10~dB SNR, the same change reduces the symbol error
rate from about $0.169$ to $0.127$ and improves tracking-CSI NMSE by
approximately $4.9$~dB. These gains arise because the prediction in
\eqref{eq:predicted_CSI_principle_enhanced} is no longer formed from a single
noisy subcarrier-wise state; instead, nearby subcarriers contribute a local
frequency-smoothed prior, which is particularly beneficial in weak-tone or
frequency-selective fading regions.

Fig.~\ref{fig:alpha_smoothing_tradeoff}(c) further shows that the same
regularization improves the sensing map. At 5~dB SNR, the TBR increases from
about $23.2$~dB at $\alpha=0$ to about $25.6$~dB at $\alpha=0.1$, giving a
map-domain contrast gain of approximately $2.4$~dB. This confirms that the
lower tracking-CSI error in Fig.~\ref{fig:alpha_smoothing_tradeoff}(b) is
converted into a cleaner range-Doppler response. The improvement is smaller at 10~dB and
nearly saturated at 15~dB because the DQPSK transition posterior is already
more concentrated; hence, smoothing can only marginally reduce the residual
background and sidelobe floor. Beyond $\alpha\approx0.1$-$0.2$, the SER, NMSE,
and TBR curves become nearly flat or slowly degrade, indicating that the
frequency-domain prior should be used as a local regularizer rather than as an
aggressive smoothing filter.

Figs.~\ref{fig:alpha_smoothing_tradeoff}(d) and
\ref{fig:alpha_smoothing_tradeoff}(e) illustrate the different roles of the
tracking and sensing fusion gains. The plotted quantities are averaged over all active carriers and post-PRS OFDM symbols, $K
    =
    \frac{1}{N_{\rm MC}F(M-1)|\mathcal K|}
    \sum_{i=1}^{N_{\rm MC}}
    \sum_{f=1}^{F}
    \sum_{m=1}^{M-1}
    \sum_{k\in\mathcal K}
    K^{(i)}_{k,m,f},$ and $G
    =
    \frac{1}{N_{\rm MC}F(M-1)|\mathcal K|}
    \sum_{i=1}^{N_{\rm MC}}
    \sum_{f=1}^{F}
    \sum_{m=1}^{M-1}
    \sum_{k\in\mathcal K}
    G^{(i)}_{k,m,f}$,
where $N_{\rm MC}$ denotes the Monte Carlo trial number, $K_{k,m,f}$ is the recursive tracking gain in
\eqref{eq:posterior_kalman_gain} and
$G_{k,m,f}$ is the sensing gain in
\eqref{eq:sensing_gain_eta_form}. 
The average tracking gain $ K$ is small and increases with SNR, from
roughly $8\times10^{-3}$ at 5~dB to about $7\times10^{-2}$ at 15~dB. 
In contrast,
the sensing gain $ G$ remains much larger, around $0.75$, $0.87$ and
$0.93$ at 5, 10 and 15~dB, respectively. This confirms the advantage of the proposed sensing fusion in Section \ref{subsec:sensing_CSI}. Namely, unreliable innovations are
suppressed, but reliable dynamic perturbations are largely retained for
radar sensing. The small increase of $ G$ with $\alpha$ is also
consistent with the posterior-reliability interpretation. In particular, a more stable
frequency-smoothed prediction makes the selected DQPSK transition more
confident, increasing $\eta_{k,m,f}$ and $G_{k,m,f}$ and allowing the sensing CSI to
admit more target-bearing innovation.

\begin{figure}[!b]
\centering
\includegraphics[width=0.95\linewidth]{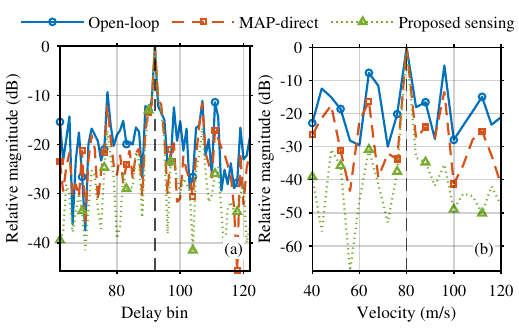}
\caption{normalized range and Doppler cuts through target T2 in Fig. \ref{fig:rdm_maps_representative_scene}:
(a) range cut and
(b) Doppler cut.
}
\label{fig:representative_rdm_cuts}
\end{figure}

\begin{figure*}[!t]
\centering
\includegraphics[width=0.9\textwidth]{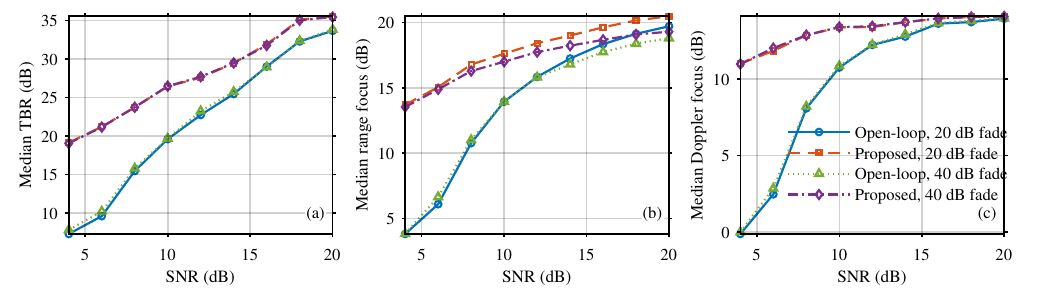}
\caption{Median range-Doppler map quality over random target scenes under controlled frequency-selective fading:
(a) TBR,
(b) range focus and
(c) Doppler focus.
The same legend applies to all subfigures.
}
\label{fig:random_scene_metrics}
\end{figure*}

\subsection{Range-Doppler Map Performance}
\label{subsec:simulation_random_scene_results}

Fig.~\ref{fig:rdm_maps_representative_scene} shows full-grid range-Doppler maps for a three-target scene at 5~dB SNR. The targets are placed at delay bins 45, 92 and 138 with velocities $-210$, 80 and 235~m/s, respectively. All three targets use the same amplitude so that the visual comparison is dominated by scheme-induced background and sidelobe behavior rather than by target-power imbalance.

The open-loop map contains the correct target locations, but they sit on a high and structured background caused by persistent differential slips. MAP-direct processing improves the map because the transition decision uses the channel prior, but hard division by $\hat X_{k,m,f}$ still admits unreliable instantaneous observations. The proposed sensing CSI produces the cleanest map. The minimum TBR over the three targets increases from $17.8$~dB for open-loop processing to $24.0$~dB for MAP-direct processing and $33.1$~dB for the proposed scheme. Thus, the proposed design provides a $15.3$~dB gain over open-loop processing and a $9.1$~dB gain over MAP-direct processing; in linear terms, the weakest target has about $34$ times higher background-normalized power than in the open-loop map and about $8$ times higher than in the MAP-direct map.
This highlights the significant sensing improvement achieved by the proposed CSI estimation scheme.

Fig.~\ref{fig:representative_rdm_cuts} examines the local range and Doppler structure around target T2 in the representative scene. The cuts are normalized to the local peak of each method for clearer comparison. In both the range and Doppler cuts, open-loop processing leaves a high and irregular local background, with several competing sidelobe peaks within 10-20~dB of the target. MAP-direct processing reduces part of this structure, but still exhibits pronounced residual sidelobes caused by hard decision errors and noisy instantaneous CSI admission.

By contrast, the proposed sensing CSI yields a much cleaner local response. Around the target delay in Fig.~\ref{fig:representative_rdm_cuts}(a), most neighboring range bins are suppressed below approximately $-20$. In the Doppler cut of Fig.~\ref{fig:representative_rdm_cuts}(b), the improvement is more pronounced. Specifically, the proposed curve keeps most off-target Doppler bins below approximately $-30$~dB, whereas open-loop and MAP-direct processing retain multiple sidelobe peaks above $-20$~dB. This reduction of the local sidelobe floor is consistent with the proposed scheme mechanism. The tracking CSI improves phase and symbol consistency, while the sensing CSI re-admits reliable innovation without allowing ambiguous DQPSK decisions to degrade CSI for sensing.

Fig.~\ref{fig:random_scene_metrics} evaluates whether the CSI improvements translate into range-Doppler map quality in random sensing scenes. For each Monte Carlo run, the target delay is drawn from 55-220 bins, the absolute target velocity from 50-240~m/s and the target amplitude from $-35$ to $-21$~dB relative to the dominant component. Two controlled frequency-selective fading depths, 20 and 40~dB, are considered. The reported values are medians over 100 independent runs. 

The most significant improvement appears in the low-SNR regime, where DQPSK transition estimation errors and CSI ambiguity are most damaging. At 4~dB SNR, the median TBR in Fig.~\ref{fig:random_scene_metrics}(a) increases from about $8.0$~dB to $19.2$~dB under the 20~dB fading and from about $7.9$~dB to $19.0$~dB under the 40~dB fading. This corresponds to an improvement of over $11$~dB, or about a $13$-fold increase in background-normalized target power. At 8~dB SNR, the proposed scheme still provides an over $7.8$~dB TBR gain, increasing the median TBR from about $15.7$~dB to about $23.7$~dB. The gap narrows at higher SNR because the open-loop DQPSK transitions become more reliable; nevertheless, the proposed scheme remains consistently superior across the full SNR range.

Figs.~\ref{fig:random_scene_metrics}(b) and~\ref{fig:random_scene_metrics}(c) further show that the gain is not limited to average target-to-background contrast. At 4~dB SNR, the median range focus improves from about $3.8$~dB for open-loop processing to about $13.6$~dB for the proposed scheme, giving nearly $10$~dB of focusing gain. At 8~dB SNR, the range-focus gain remains about $5.3$~dB. The Doppler-focus improvement is also pronounced. At 4~dB SNR, the open-loop map has nearly zero Doppler focus, whereas the proposed scheme achieves about $10.8$~dB. At 8~dB SNR, the Doppler-focus gain remains around $4$~dB. These results indicate that open-loop errors do not merely raise the background floor; they also spread target energy across delay and Doppler cells. By contrast, the proposed sensing CSI preserves target-bearing innovation while suppressing unreliable decision-induced perturbations, giving a more compact range-Doppler peak in both dimensions.

Though the 20 and 40~dB fading curves are close for both schemes, the proposed 
scheme consistently outperforms the open loop scheme over the whole SNR range. This highlights the robustness of the proposed design under random target delays, velocities, amplitudes, clutter and noise realizations.

\subsection{Experimental Results}
\label{subsec:experimental_results}

\begin{figure}[!t]
\centering
\includegraphics[width=0.9\linewidth]{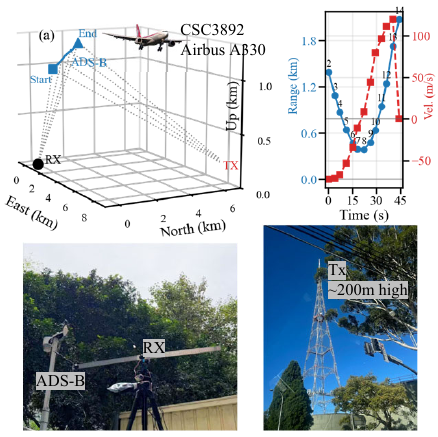}
\caption{Experimental bistatic geometry and ADS-B-derived reference quantities for the selected aircraft pass:
(a) three-dimensional geometry of the DAB+ transmitter, receiver and aircraft trajectory and
(b) ADS-B-derived bistatic range and bistatic velocity over the observation interval. The numbered markers indicate the processed DAB+ captures.}
\label{fig:experiment_diagram}
\end{figure}

We further validate the feasibility of the proposed design using passive commercial-aircraft sensing experiments in Sydney, Australia. Fig.~\ref{fig:experiment_diagram} illustrates the experimental scenario. The illuminator is the commercial DAB+ transmitter at Artarmon\footnote{Details of the transmitter can be found here: \url{https://web.acma.gov.au/rrl/assignment_search.lookup?pEFL_ID=5613134}.}, which operates with vertical polarization and a licensed maximum ERP of 50~kW. The receiver is located near Homebush train station, approximately 10~km from the Artarmon transmitter. We use a low-cost 8-bit RTL-SDR receiver, a single antenna and a Raspberry Pi for data capture at the 202.928~MHz carrier frequency. A second RTL-SDR connected to the same Raspberry Pi is used to record ADS-B messages\footnote{ADS-B provides periodically broadcast aircraft position and velocity information derived from onboard navigation systems; open source tools and tutorials are widely avaialble; see e.g., \url{https://github.com/littleBane/adsb-rpi}.} during the DAB+ acquisition. 
Since the focus of this paper is reliable DAB+ CSI estimation, the experiment is intended as a feasibility demonstration rather than a complete passive-radar measurement campaign. Below we present one representative aircraft pass associated with an Airbus A330.

\begin{figure}[!t]
\centering
\includegraphics[width=0.9\linewidth]{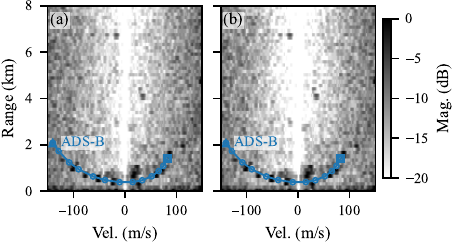}
\caption{Experimental stacked range-Doppler responses over multiple DAB+ captures:
(a) open-loop differential processing and
(b) proposed scheme. The overlaid ADS-B trace is obtained from the geometry in Fig.~\ref{fig:experiment_diagram}.}
\label{fig:stacked_rdm}
\end{figure}

\begin{figure}[!t]
\centering
\includegraphics[width=0.9\linewidth]{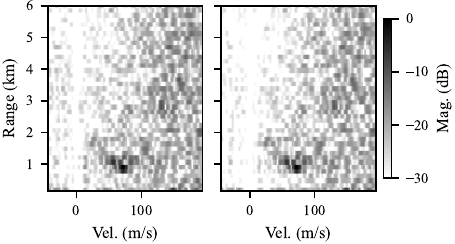}
\caption{Representative experimental single-capture range-Doppler maps:
(a) open-loop differential processing and
(b) proposed scheme.}
\label{fig:single_rdm}
\end{figure}

\begin{figure}[!t]
\centering
\includegraphics[width=\linewidth]{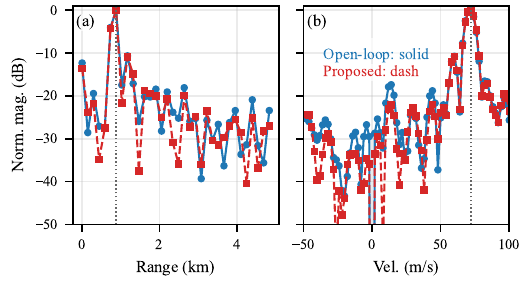}
\vspace{-8mm}
\caption{Local cuts of the representative experimental range-Doppler map in Fig.~\ref{fig:single_rdm}:
(a) range cut and
(b) Doppler cut.}
\label{fig:rdm_cuts_experiment}
\end{figure}

Fig.~\ref{fig:stacked_rdm} compares the accumulated range-Doppler responses obtained from approximately 15 DAB+ captures along the selected aircraft pass. The range-Doppler map from each capture is first individually normalized and then incoherently added up for visual checking of the flying pass. The ADS-B-derived trajectory is overlaid as a true reference. Both methods show energy near the expected aircraft region, confirming that the single-stream DAB+ CSI sequence contains target-induced dynamic components. However, the open-loop response is more diffuse and is accompanied by stronger background structure. In contrast, the proposed sensing-CSI output produces a cleaner accumulated response along the ADS-B-consistent trajectory, demonstrating the practical benefit of posterior-aware differential CSI tracking. Fig.~\ref{fig:single_rdm} presents a representative single-capture range-Doppler map. Compared with open-loop processing, the proposed method suppresses part of the diffuse background and yields a more localized response around the ADS-B-consistent range-velocity region. The corresponding local cuts in Fig.~\ref{fig:rdm_cuts_experiment} provide a clearer comparison. The proposed method has lower sidelobe levels in both range and Doppler dimensions.

Overall, the experimental results show that the proposed CSI estimation framework can be applied to real-life DAB+ passive radar and can provide a cleaner range-Doppler representation than open-loop processing. The improvement is less pronounced than in the controlled simulations, which is expected. The measurement used an 8-bit low-cost SDR front end, so quantization, limited dynamic range and possible front-end compression may affect the recovered CSI in the presence of a strong direct-path component from the 50~kW transmitter. The receiving antenna was horizontally polarized to reduce the vertically polarized direct-path component, but this provides only partial suppression. In addition, the Sydney DAB+ environment contains other transmitters and repeaters and the receiver location is close to major roads, which can introduce additional multipath, interference and moving clutter. A systematic experimental study with calibrated front ends, controlled antenna orientation, multi-pass aircraft data and more complete clutter suppression is left for future work.

\section{Conclusion}
\label{sec:conclusion}

This paper proposes a posterior-probability-aware differential CSI tracking framework for DAB+-based passive radar 
to address the challenge of error propagation in estimating CSI from differentially modulated symbols in a DAB+ frame. 
The proposed scheme combines frequency-smoothed CSI prediction, prediction-aided MAP transition detection, posterior-aware LMMSE CSI tracking, and a separate reliability-informed sensing-CSI output that preserves target-bearing innovation.
Theoretical analysis illustrates the transition-SNR advantage over open-loop differential processing, the bias-variance role of frequency smoothing, and the CSI-error suppression achieved by posterior-aware fusion. Simulation results 
show that at a low post-FFT SNR of 5~dB, 
the proposed method improves the NMSE of CSI estimation and the target-to-background ratio in range doppler maps, each by over 15~dB, as compared with prior art. Experimental results also demonstrate improved range-Doppler maps of sensing a commercial aircraft in Sydney, Australia. 

\appendix

\subsection{Frequency Smoothing and Local Bias-Variance Behavior}
\label{app:frequency_prediction}

This appendix analyzes the prediction model in \eqref{eq:channel_prior_principled_enhanced} and the smoothing operator in \eqref{eq:smoothing_operator_principled_enhanced}. For an interior active tone with two neighbors, write the previous posterior estimate as
\begin{equation}
  \hat H_{k,m-1,f}=H_{k,m-1,f}+E_{k,m-1,f},
  \label{eq:app_prev_error_model}
\end{equation}
where $E_{k,m-1,f}$ denotes the previous CFR estimation error. Let $\Delta H_{k,m,f}=H_{k,m,f}-H_{k,m-1,f}$
be the one-step temporal innovation. Based on \eqref{eq:smoothing_operator_principled_enhanced} we have
\begin{align}
  &\tilde H_{k,m,f}-H_{k,m,f}\label{eq:app_smoothing_error_decomposition}\\
  & =
  \frac{\alpha}{2}
  \big(H_{k-1,m-1,f}-2H_{k,m-1,f}+H_{k+1,m-1,f}\big)-\Delta H_{k,m,f}
  \nonumber\\
  &
  +(1-\alpha)E_{k,m-1,f}
  +\frac{\alpha}{2}\big(E_{k-1,m-1,f}+E_{k+1,m-1,f}\big).
  \nonumber
\end{align}
The first term is a frequency-curvature bias, the second term is temporal process innovation, and the remaining terms are filtered estimation errors. If the neighboring estimation errors are locally uncorrelated with common variance $\sigma_E^2$, the filtered-error variance is
\begin{equation}
  \sigma_{E,{\rm filt}}^{2}
  =
  \rho_{\alpha}\sigma_E^2,
  ~
  \rho_{\alpha}=(1-\alpha)^2+\frac{\alpha^2}{2}.
  \label{eq:app_smoothing_variance_factor}
\end{equation}
Thus mild smoothing reduces subcarrier-wise estimation-noise variance before transition detection, while the curvature term in \eqref{eq:app_smoothing_error_decomposition} prevents using a large $\alpha$.

The physical origin of local frequency correlation follows from \eqref{eq:channel_expansion}. The difference $H_{k+1,m,f}-H_{k,m,f}
  =
  \sum_{p=0}^{P-1} a_p[m,f]
  e^{-\jmathi2\pi k\Delta f\tau_p[m,f]}
  \left(e^{-\jmathi2\pi\Delta f\tau_p[m,f]}-1\right),$
yields
\begin{equation}
  |H_{k+1,m,f}-H_{k,m,f}|
  \leq
  2\pi\Delta f
  \sum_{p=0}^{P-1}|a_p[m,f]|\tau_p[m,f].
  \label{eq:app_adjacent_bound}
\end{equation}
Since $\Delta f=1$~kHz in DAB Mode I, low-delay direct and clutter components are strongly correlated across adjacent subcarriers, while long-delay components contribute more curvature. This is why \eqref{eq:smoothing_operator_principled_enhanced} is used only as a weak local predictor rather than as a global smoothing constraint.

\subsection{MAP Detector and Posterior Hypothesis Probabilities}
\label{app:map_detector}

For a candidate transition $q\in\mathcal Q$, the likelihood in \eqref{eq:transition_likelihood_principled_enhanced} is
\begin{equation}
  p(Y_{k,m,f}|q)
  =
  \frac{1}{\pi\sigma_{\me,k,m,f}^{2}}
  \exp\!\left(
  -\frac{r_{k,m,f}(q)}{\sigma_{\me,k,m,f}^{2}}
  \right),
  \label{eq:app_transition_likelihood}
\end{equation}
where $r_{k,m,f}(q)$ is defined in \eqref{eq:transition_residual_def}. With equal transition priors, maximizing $p(q|Y_{k,m,f})$ is equivalent to maximizing \eqref{eq:app_transition_likelihood}, which is equivalent to minimizing $r_{k,m,f}(q)$. This gives \eqref{eq:map_transition_principled_enhanced}.
Normalizing the four likelihoods in \eqref{eq:app_transition_likelihood} gives the posterior probability
in \eqref{eq:dqpsk_posterior_probability_final}.

\subsection{Observation Variance Induced by DQPSK Ambiguity}
\label{app:observation_variance}

Using \eqref{eq:dqpsk_transition_model_enhanced} and \eqref{eq:symbol_update_principled_enhanced}, the true and reconstructed symbols satisfy
\begin{equation}
  X_{k,m,f}=X_{k,m-1,f}q_{k,m,f},
  ~
  \hat X_{k,m,f}=\hat X_{k,m-1,f}\hat q_{k,m,f}.
  \label{eq:app_true_recon_symbols}
\end{equation}
Conditioned on the previous symbol being correctly reconstructed, or equivalently with its uncertainty absorbed into \eqref{eq:transition_residual_variance}, the  observation in \eqref{eq:instantaneous_CSI_observation_enhanced} becomes
\begin{align}
  &Z_{k,m,f}
  =
  H_{k,m,f}\frac{q_{k,m,f}}{\hat q_{k,m,f}}
  +\frac{W_{k,m,f}}{\hat X_{k,m,f}}
  \nonumber\\
  &=
  H_{k,m,f}
  +\frac{W_{k,m,f}}{\hat X_{k,m,f}}
  +H_{k,m,f}\left(\frac{q_{k,m,f}}{\hat q_{k,m,f}}-1\right).
  \label{eq:app_observation_error_decomposition}
\end{align}
Because $|\hat X_{k,m,f}|=1$, the additive-noise term in \eqref{eq:app_observation_error_decomposition} has variance $\sigma_0^2$. Conditioning on $Y_{k,m,f}$ and using the posterior probabilities in \eqref{eq:dqpsk_posterior_probability_final}, the symbol-ambiguity-induced error power is 
\begin{equation}
  |H_{k,m,f}|^2
  \sum_{q\in\mathcal Q}
  \pi_{k,m,f}(q)
  \left|\frac{q}{\hat q_{k,m,f}}-1\right|^2 .
  \label{eq:app_symbol_ambiguity_power}
\end{equation}
Replacing the unknown $|H_{k,m,f}|^2$ by the predicted power $|\tilde H_{k,m,f}|^2$ gives the implementable variance estimation in \eqref{eq:posterior_observation_variance_final}.

\subsection{LMMSE Fusion Gain}
\label{app:lmmse_fusion}

Consider the conditional scalar model
\begin{equation}
  \tilde H_{k,m,f}=H_{k,m,f}+E^{\rm p}_{k,m,f},
  ~
  Z_{k,m,f}=H_{k,m,f}+E^{\rm z}_{k,m,f}, \nonumber
\end{equation}
where $E^{\rm p}_{k,m,f}$ and $E^{\rm z}_{k,m,f}$ are conditionally zero-mean, uncorrelated, and have variances $\sigma_{\sfp,k,m,f}^{2}$ and $\sigma_{\mz,k,m,f}^{2}$, respectively. For the linear update
\begin{equation}
  \hat H_{k,m,f}(K)
  =
  \tilde H_{k,m,f}+K\left(Z_{k,m,f}-\tilde H_{k,m,f}\right),
  \label{eq:app_linear_update}
\end{equation}
its conditional MSE is
\begin{equation}
  \mathcal E(K)
  =
  (1-K)^2\sigma_{\sfp,k,m,f}^{2}
  +K^2\sigma_{\mz,k,m,f}^{2}.
  \label{eq:app_fusion_mse}
\end{equation}
Differentiating \eqref{eq:app_fusion_mse} with respect to $K$ and setting the derivative to zero gives the LMMSE fusion gain in 
\eqref{eq:posterior_kalman_gain}, i.e., $K^{\star}_{k,m,f}$. The corresponding posterior CFR error variance satisfies
\begin{equation}
  \sigma_{{\rm post},k,m,f}^{2}
  =
  \frac{\sigma_{\sfp,k,m,f}^{2}\sigma_{\mz,k,m,f}^{2}}
  {\sigma_{\sfp,k,m,f}^{2}+\sigma_{\mz,k,m,f}^{2}}
  \leq
  \min\{\sigma_{\sfp,k,m,f}^{2},\sigma_{\mz,k,m,f}^{2}\}. \nonumber
\end{equation}

\bibliographystyle{IEEEtran}
\bibliography{ref3}

\end{document}